\documentclass[prd,superscriptaddress, longbibliography, showpacs, showkeys,a4paper,notitlepage,reprint]{revtex4-1}
\usepackage{amssymb,amsmath,bm}
\usepackage{graphicx}
\usepackage{hyperref}
\hypersetup{
  pdftitle = {Is there a map between Galilean relativity and special relativity?},
 pdfauthor = {A. Shariati and N. Jafari},
pdfsubject = {Relativity, Lorentz Group},
pdfkeywords = {Relativity, Lorentz Group},
pdfpagemode = {UseOutlines},
bookmarks,
bookmarksopen,
pdfstartview = FitH,
colorlinks,
linkcolor = blue,
citecolor = blue,
urlcolor = blue}
\newcommand{\com}[2]{\left[#1,#2\right]}

\newcommand{\Mink}{\mathbb{M}}
\newcommand{\Gal}{\mathbb{G}}
\newcommand{\Poincare}[1][{4}]{\mathop{\text{Poi}}(#1)}
\newcommand{\Lorentz}[1][{4}]{\mathop{\text{Lor}}(#1)}
\newcommand{\Newton}[1][{4}]{\mathop{\text{Newt}}(#1)}
\newcommand{\Galgroup}[1][{4}]{\mathop{\text{Gal}}(#1)}
\newcommand{\pder}[2]{\frac{\partial#1}{\partial#2}}

\newcommand{\Real}{\mathbb{R}}

\begin{document}
\title{Is there a map between Galilean relativity and special relativity?}
\author{A. Shariati}
\date{\today}
\affiliation{Physics Group, Faculty of Sciences, Alzahra University, Tehran, 19938, Iran.}
\email{shariati@mailaps.org}
\author{N. Jafari}
\affiliation{Science and Research Branch, Islamic Azad University, 
 Hesarak, Tehran, 14778, Iran.}
\email{nosrat.jafari@gmail.com}  
\begin{abstract}
Mandanici in \cite{Mandanici2014} has provided a map which he claims to be
a two way map between Galilean relativity and special relativity.  We argue that
this map is simply a curvilinear coordinate system on a subset of the two-dimensional
Minkowski space-time, and is not a two way map between 1+1 dimensional 
Galilean relativity and 1+1 dimensional special relativity. 
\end{abstract}
\maketitle

\section{Introduction}
In concise mathematical terms, special relativity means covariance under the action
of the Poincar\'e groupو while Galilean relativity means covariance under the action of
the full Galilean group.   
The space-time of special relativity is the $3+1$-dimensional Minkowski space-time,
$\Mink^4\sim\Real^4$, which is a manifold on which a semi-Riemannian metric
$ds^2 =c^2\, dt^2 - dx^2 - dy^2 - dz^2$ is defined 
\cite[see for example][p. 118]{HawkingEllis}.  The space-time of the Galilean 
relativity is the $3+1$-dimensional Galilei space-time, $\Gal^4\sim\Real^4$, which is an affine
space, with an \emph{absolute} time function \cite[see][pp. 3-6]{Arnold1989}.

The symmetry group of the $(n+1)$-dimensional Minkowski
space-time, $\Mink^{n+1}$, is the Poincar\'e group $\Poincare[n+1]$, consisting of 
translations, rotations, and Lorentz transformations.  Rotations and Lorentz 
transformations form a group, $\text{O}(n,1)$.  The subgroup $\text{SO}^\uparrow(n,1)$, consisting of special, orthochronous transformations is denoted by $\Lorentz[n+1]$.
The symmetry group of the $(n+1)$-dimensional Galilean space-time, $\Gal^{n+1}$, is the
`Newtonian' group $\Newton[n+1]$, consisting of translations, rotations, and Galilean
transformations.  The special orthochronous sub-group of the rotations and the
Galilean transformations is the Galilean group, which we denote by $\Galgroup[n+1]$.

As groups, $\Lorentz[2]$ and $\Galgroup[2]$ are both isomorphic to $(\Real, +)$.  
In higher dimensions the situation is different.  For $n > 1$ the groups
$\Lorentz[n+1]$ and $\Galgroup[n+1]$ are both $\frac 1 2 n (n+1)$-dimensional, but
they are not homomorphic, because their Lie algebras are not homomorphic.  
The groups $\Poincare[n+1]$ and $\Lorentz[n+1]$ depend on 
a parameter, $c$.  It can be shown that the $c\to\infty$ limit of these groups are,
respectively $\Newton[n+1]$ and $\Galgroup[n+1]$---the In\"on\"u-Wigner contraction
\cite{Inonu-Wigner}.
It is also important to notice that the symmetry algebra of a non-relativistic quantum
system, is not the Lie algebra of $\Newton[n+1]$, but a larger \emph{central extension} of
that---for a textbook introduction to In\"on\"u-Wigner contraction and the 
emergence of this central extension see \cite[pp. 61-62]
{WeinbergFields1}.

\section{The Mandanici's Claim}
In a recent article, G.~Mandanici introduced a map, which he claims to be 
a two way map between the special and the Galilean relativities \cite{Mandanici2014}.

\paragraph*{The Mandanici map.}
Consider the two-dimensional Minkowski space-time $\Mink^2$ with Cartesian coordinates 
$(T,X)$ and metric
\begin{math} ds^2 = c^2\, dT^2 - dX^2 \end{math}. Now consider curvilinear coordinates
$(t, x)$ defined by the following formulas.
\begin{align} \label{eq:1a}
X & = \alpha\, c\, t\, \sinh \left( \frac{x}{c\, t} \right) +
\beta \, c \, t\, \cosh \left(\frac{x}{c\, t} \right), \\
\label{eq:1b}
T & = \alpha\, t \, \cosh\left( \frac{x}{c\, t} \right) + 
\beta\, t\, \sinh\left(\frac{x}{c\, t}\right),
\end{align}
where $\alpha$ and $\beta$ are dimensionless parameters.  
Calculating the Jacobian is straightforward:
\begin{equation}
J = \pder{T}{t} \cdot \pder{X}{x} - \pder{T}{x} \cdot \pder{X}{t}
= \alpha^2 - \beta^2,
\end{equation}
from which it follows that this transformation is invertible, only if
$\alpha \neq \beta$.  Consider the special case of $\alpha = 1$, $\beta = 0$,
which means 
\begin{equation} \label{eq:2}
 X  =  c\, t\, \sinh\left(\frac{x}{c\,t}\right), \quad
 T  =  t\, \cosh\left(\frac{x}{c\,t}\right). \end{equation}
Obviously 
\begin{math} c^2\, T^2 - X^2 = c^2\, t^2 \end{math}
and \begin{math} X/c\, T = \tanh(x/c\, t) \end{math}.  From these two
equations, it is obvious that this transformation ($\alpha = 1$, $\beta = 0$)
is defined only for the region $c^2 \, T^2 \geq X^2$ of the 2D Minkowski space-time.

On $\Mink^2$, the vector field
\begin{math}
K = c^{-1}\, X \, \partial_T + c\, T \, \partial_X
\end{math}
is a Killing vector field, which generates Lorentz boosts.
(A Killing vector field is a vector field which generates
an isometry of the space-time, that is, a symmetry of the space-time.)
Now let's see what is the vector field $\partial_x$.  It is easy to see that
\begin{align}
 \pder{}{x} & = \pder{T}{x}\, \pder{}{T} + \pder{X}{x}\, \pder{}{X}
\cr & = \frac{1}{c} \sinh\left(\frac{x}{c\, t}\right) \pder{}{T}
+ \cosh\left(\frac{x}{c\, t}\right)\pder{}{X} \cr
\partial_x & = \frac{1}{c\, t} \left( c^{-1}\, X\, \partial_T + c\, T\, \partial_X \right)
\end{align}
from which we see that
\begin{equation}
c\, t\, \partial_x = c^{-1}\, X\, \partial_T + c\, T\, \partial_X = K.
\end{equation}
In words: The vector filed $K$, in terms of the curvilinear coordinates $(t,x)$ reads
\begin{math}
c\, t\, \partial_x
\end{math}.
The transformation this vector field generates, in terms of the curvilinear coordinates
$(t,x)$, is
\begin{math}
(t, x) \mapsto (t, x + \theta\, c\, t)
\end{math},
where $\theta$ is the dimensionless parameter of the flow.  Defining $v : = c\, \theta$,
this is the famous Galilean transformation.	
This does not mean that transformation (\ref{eq:2}) transforms the 2D Miknowski space-time
into the 2D Galilean space-time!
What is given by the map (\ref{eq:2}), is a change of coordinates in a subset of 
$\Mink^2$.  On this subset---the interior of the light-cone---the Killing vector field 
$K$, in terms of the curvilinear coordinates $(t,x)$ has the form $K = c\, t\, \partial_x$, 
which is the generating vector field of the
Galilean transformations.  However, this does not mean that we have found a map from the
Newtonian group $\Newton[2]$ to the Poincar\'e group $\Poincare[2]$?  

Remembering that both
$\Galgroup[2]$ and $\Lorentz[2]$ are isomorphic to $(\Real, +)$, this is not strange.
In fact, according to the straightening-out theorem,
any smooth vector field, in the neighborhood of a point at which the vector field is non-zero,
could be written as $\partial_1$, for a suitably chosen coordinate system
$(x^1, x^2)$ \cite[see for example][p. 177]{Curtis-Miller1985}, which means that
if we wish, we could find a map such that the Lorentz boosts in $\Mink^2$ have the
form of translations---locally, of course.
However, besides $K$, the Minkowski space-time $\Mink^2$ has two other Killing vector fields,  
$\partial_T$, and $\partial_X$, which are generators of translations in time and space.
Transformation (\ref{eq:2}) does not map these two vector fields to 
$\partial_t$ and $\partial_x$.  

\paragraph*{Minor comments.}
The notation of Mandanici's article is misleading.  The infinitesimal action
of a vector field on the manifold is $x^\mu \to x^\mu + \delta x^\mu$.  Writing it as
$\Delta x^\mu \to \Delta x^\mu + \delta \Delta x^\mu$ caused the author to consider it,
not as a flow on the manifold, but as a map of the tangent space (the velocity space), 
which is not correct---see equations (1) and (2) of \cite{Mandanici2014}.
Due to this confusion, the author confused $X/c\, T$ and $x/c\, t$, with velocities (in
his notation with $v_{E}$ and $v_{G}$).  But these ratios are not velocities---velocities 
of what object should they be?  The relation $X/c\, T = \tanh \left( x/c\, t\right)$, 
which the author writes as $v_{E} = \tanh v_{G}$, simply shows that the map (\ref{eq:2})
is defined on the closure of the interior of the light cone, \textit{i.e.},
for $c^2\, T^2 - X^2 \geq 0$.
\par $\Newton[n+1]$ is not isomorphic
to $\Poincare[n+1]$, because, as Lie groups, the corresponding Lie algebras are
not homomorphic.  Therefore, from the onset, we know that one cannot find a two way
map sending a subset of $\Mink^{n+1}$ to a subset of $\Gal^{n+1}$, mapping the
corresponding structures.
\vspace*{1cm}

\section{Different relativities?}
Almost two decades ago, a set of theories, named doubly special relativity (DSR) theories
emerged.  It is claimed that such theories incorporate an invariant length or energy 
scale in the special relativity, and since special relativity has already an invariant 
velocity scale, these theories are named `doubly' special relativities.  
Some authors, including ourselves, believe that \emph{some}
of these theories are reformulations of the special theory of relativity in curvilinear 
coordinate systems \cite{JS04, jafari:462, JS11}.   
Specifically, the Fock-Lorentz and the Magueijo-Smolin DSRs are `projective' representations
of the Lorentz group see \cite{JS11}.
We should emphasize that `projective' actions
of the Lorentz group had been studied by E. H. Kerner in 1976 
\cite{kerner1976a, kerner1976b}, but his works are not properly
cited in the literature. We did not know of his works, until our paper was published.

Briefly, the point is that, a coordinate system is not important.   What \emph{is} important,
is the physical and geometric relations between the objects.    To clarify this point, let's 
return to the case of Galilean \emph{vs.} special relativity.  In $1+1$ dimensions,
Galilean relativity means invariance under 
$H = \partial_t$, $P =\partial_x$, and $K = t\,\partial_x$;
satisfying $\com{H}{P} = 0$, $\com{H}{K} = P$, $\com{K}{P} = 0$.
(The commutation relation $\com{K}{P} = 0$ is valid for transformations
of the space-time events.  For the state space of a particle of mass $m$, 
this should be replaced by $\com{K}{P} = -m$.  This algebra, is a central extension of the 
one acting on the space-time events.)
Einstein's relativity, in $1+1$ dimensions, means invariance under 
$H = \partial_T$, $P = \partial_X$, and
$K = c^{-1}\, X\, \partial_T + c\, T\, \partial_X$, satisfying
$\com{H}{P} = 0$, $\com{H}{K} = c\, P$, $\com{P}{K} = c^{-1}\, H$.
These two Lie algebras are not homomorphic, so that one cannot obtain one from the other
by a change of coordinates.  
Therefore, Galilei relativity and special relativity are distinct theories---one cannot find
a two way map between them, and their physical content is not the same.

The case of some doubly special relativities, such as the Fock-Lorentz and
the Magueijo-Smolin theories is the reverse: It is possible to find coordinate changes, which
transform them to special relativity \cite{JS04, jafari:462, JS11}, and that, we believe,
means equivalence.

Some authors believe that two different sets of transformations could describe different
physical theories, even if they are mathematically isomorphic \cite[see \S 3.1 of][]{AC11}.
Mandanici in \cite{Mandanici2014} has tried to find a two way map between the Galilean 
relativity and the special relativity, providing an example showing that two 
physically non-equivalent theories could be mathematically isomorphic.   Our points are 
as follows.
\begin{enumerate}
\item The map Mandanici has provided is not a two way map between the Galilean relativity and
the special relativity.  This map only changes the form of 
$c^{-1}\, X\, \partial_T + c\, T\, \partial_X$ to $c\, t\, \partial_x$, and this is not
surprising---since both generate a 1D group of transformations, isomorphic to $(\Real, +)$.
\item Because the Lie algebra of the full Galilean group and the Poincar\'e group are not 
homomorphic, one cannot find a coordinate system on $\Mink^4$ such that the Killing vector
fields are transformed to the vector fields generating the full Galilean group of
transformations.
\end{enumerate}

\begin{acknowledgments}
 This work was supported by the research council of Alzahra University.
\end{acknowledgments}

\bibliography{JS2bibtex}
\include{SJ-14a.bbl}
\end{document}